\documentstyle[12pt,epsf]{article}
\begin{document}

\hfill {WM-97-109}

\def\tcww{$t\ \rightarrow\ c\ W^+W^-$}
\def\wwtc{$W^+W^- \rightarrow\ c\bar{t}+t\bar{c}$}

\hfill {June, 1997}

\vskip 1in 
{

   \bigskip \centerline{\bf 
$t\to c WW$ and $WW\to \bar t c+t \bar c$ in Extended Models
} }
 \vskip .8in
\centerline{David Atwood}\bigskip
\centerline{\it Theory Group, Thomas Jefferson National Accelerator
Facility, Newport News, VA  23606, USA}
 \bigskip
\centerline{Marc
Sher } 
\bigskip
\centerline {\it Physics Department, College of William and
Mary, Williamsburg, VA 23187, USA}

\vskip 1in
 
{\narrower\narrower   Jenkins has pointed out that the process
$t\rightarrow cW^+W^-$ is GIM suppressed in the standard model. In
this note, we calculate the branching ratio for a wide range of
models, in which the decay occurs at tree level through exchange of a
scalar, fermion or vector.  In the case of scalar exchange, a
scalar mass between $2m_W$ and $200$ GeV leads to a resonant
enhancement, giving a branching ratio as high as a few tenths of a 
percent.  We then
note that all of these models will also allow $W^+W^-\rightarrow
\bar{t}c+t\bar{c}$, and we calculate the single-top/single-charm
production rate at the LHC.  
The rates are such that
the background from single-top/single-bottom
production will probably swamp the signal.}

\newpage

\section{Introduction}

It has now been established that the mass of the top quark is well
above $163$ GeV.  This makes the decay \tcww\ kinematically
accessible.  The branching ratio for this decay in the Standard Model
was studied by Jenkins\cite{jenkins}.   The diagram is given in Figure
1, where the internal fermion can be any of the charge $-1/3$ quarks. 
Jenkins  showed that the rate is extremely small.  There are two
reasons for this.  The first, of course, is phase space.  The second
is a GIM cancellation, which makes the resulting rate proportional to
the square of the $b$-quark mass.  Jenkins noted that this second
suppression might be absent in other models, and that in these models
the rate might be measurable.

In this letter, we first look at the decay \tcww\ in a very generic
set of models, in which the exchanged particle can be either a scalar,
a fermion or a vector boson.  We then note that in  any model in
which the decay
\tcww\ occurs, one will also have the process \wwtc.  This process
will  not be phase-space suppressed, and could thus be much
larger.  We will calculate the rate for $p\ p\ \rightarrow W^+W^- 
\rightarrow\  c\bar{t}+t\bar{c}$ for the same set of
models.\footnote{The related process
$e^+e^-\rightarrow\ c\bar{t}+t\bar{c}+\nu\bar{\nu}$ as well as the
rate for \tcww\ , in the particular case of scalar exchange, has been
calculated by Bar-Shalom et al.\cite{barsh}}

In section 2, we discuss the generic set of models that will be
considered.  The rate for \tcww\ in these models will be calculated
in section 3, and the rate for $p\ p\ \rightarrow W^+W^- \rightarrow\
c\bar{t}+t\bar{c}$ will be determined in section 4.   Section 5
contains a discussion of the results and conclusions.

\section{Models}

The most general set of operators involving a $t$, $c$, $W^+$ and
$W^-$ will have 36 coefficients, each an arbitrary function of the
appropriate Mandelstam variables.  Rather than try to deal with this
large set, we will restrict the discussion to models in which the
\tcww\ and \wwtc\ processes can occur at tree-level.  This seems
reasonable; the rates are already very small, and in models in  which
these processes arise from loops and higher-dimensional operators,
one would expect the rates to be even smaller.  Thus, for
example, the coupling of a fermion to the $W$-bosons will be
taken to be the most general combination of $V$ and $A$ but not $T$,
since tensor couplings generally arise from loops.  Other than that
restriction, the models will be completely general; the exchanged
particle can either be a scalar, fermion or vector boson.

One of the simplest extensions of the standard model is the two-Higgs
model, in which an additional scalar doublet is included.  In general,
such an extension will have tree-level flavor-changing neutral
currents.  One can avoid such currents\cite{gw} by either coupling all
of fermions to one doublet, or by coupling the charge $2/3$ quarks to
one doublet and the charge $-1/3$ quarks to the other.  This can be
done via a discrete symmetry.  However, it has been pointed
out\cite{cs} that  there is no phenomenological need to do this, as
long as the couplings of the scalars to the light quarks is
small.  From an analysis of mass matrices, Cheng and Sher\cite{cs}
argued that the most natural value for the flavor-changing couplings
of the neutral scalars to fermions is the geometric mean of the
Yukawa couplings of the fermions, i.e. the top-charm-scalar coupling
should be of
$O(\sqrt{(g_Y)_t(g_Y)_c})$, where $g_Y$ is the conventional Yukawa
coupling.  From this ansatz, one finds that the lower bound on the
exchanged scalar mass is fairly weak, of $O(100-300)$ GeV (the precise
bound depends on the amount of fine-tuning one is willing to
tolerate).  We will scale the top-charm-scalar coupling constant by
this factor.

The diagram for scalar exchange is given in Figure 2.  We have left
the coupling arbitrary.  As discussed above, the most natural value
for $a$ and $b$ is $O(1)$, although in principle they could be larger
(perturbation theory would become questionable if they were much
larger than $O(10)$).  Since the Higgs-$W^+$-$W^-$ coupling depends
on the Higgs vacuum expectation value, one expects $C_w$ to be a
ratio of vacuum values, i.e. of $O(1)$.

The diagram for fermion exchange is given in Figure 1.  In the
standard model, we expect $\alpha=\alpha^\prime=-1$, and the
contributions of the d,s and b quarks all cancel in the massless
limit, leading to a result that the diagram is GIM suppressed.
If there is a fourth generation, however, the internal fermion must
be heavy. 

The diagram for vector exchange is given in Figure 3.    Here, we
expect the contribution to be small.  In order to interact with two
$W$'s, the neutral vector boson must mix with the $Z$-boson.  Since
the properties of the $Z$ are in stunning agreement with theoretical
expectations, any such mixing must be very small.  In a wide range of
models\cite{zmix}, the mixing must be much less than $0.01$, and thus
we expect the effective coupling, $g_1$, to be less than one percent of
the weak gauge coupling.  The coupling $g_2$ can be large, but the
mass of the vector boson must be greater than $170$ GeV to avoid
dominating top decays.

We now turn to the calculation of \tcww\ in these models.

\section{\tcww}

The top quark mass is known to be between (approximately) $170$ and
$180$ GeV.  In order to get the largest plausible value for the
branching ratio for \tcww, we will take it to be $180$ GeV.  The
differential decay rate is given by
\begin{equation}
{d\Gamma\over dE_+dE_-}={|{\cal M}|^2\over 64\pi^3M_t}
\end{equation}
where $E_\pm$ are the energies of the emitted $W^\pm$.  Dividing by
the standard model decay rate of the top quark, which to within a few
percent is given by $G_FM^3_t/(8\pi\sqrt{2})$, gives the branching
ratio, after integrating over $E_\pm$.  For each of the models
discussed above, we will give the square of the matrix element, and
plot the resulting branching ratio.  The polarizations of the final
state
$W$'s will be summed over, and the charmed quark mass will be set
equal to zero, except when it appears as a parameter in the scalar
Yukawa coupling.

For the model with scalar exchange, shown in Figure 2, the matrix
element squared is given by
\def\beq{\begin{equation}}
\def\eeq{\end{equation}}
\beq
|{\cal M}|^2={g^4M_tM_cC^2_w(a^2+b^2)\over
(s-M^2_s)^2+\Gamma^2_SM^2_s} p_t\cdot p_c\bigg[\left({s\over
2m^2_W}-1\right)^2+2\bigg]
\eeq
where $p_t\cdot p_c\ =M_t(M_t-E_+-E_-)$ and $s=M_t(2E_++2E_-M_t)$.
For the scalar width, we use the expression of Atwood et
al.\cite{atw}, and include the effects of $S\rightarrow WW^*$ (which
has a small but noticeable effect for $M_s$ very near $2m_W$)
Inserting into the expression for the branching ratio gives the result
in Figure 4.  There, we have taken
$C_w^2(a^2+b^2)=1$.

One can see the resonance where the scalar mass is between $2m_W$ and
$M_t$.  In this region, the scalar can be on-shell, leading to a
much larger rate.  In this case, the decay would be clearly
measureable.  Outside of the resonance region, however, the rate
falls very rapidly.\footnote{This resonance region was also noted
in the work of Bar-Shalom, et al.\cite{barsh}.  Our result for this
decay is in agreement with theirs.}

The diagram for fermion exchange is given in Figure 1.  The minimum
value for the fermion mass,  $M_f$, is $130$ GeV.  If we define
$a=1+\alpha\alpha';\ b=\alpha+\alpha';\ a'=1-\alpha\alpha'$ and
$b'=\alpha-\alpha'$, then the matrix element squared can be written
as 
\beq
|{\cal M}|^2={g_1^2g_2^2|V_{cf}|^2|V_{ft}|^2\over
4(k^2-M_f^2)^2}\left(M_t^2(a^2+b^2)T_1 +M_f^2(a^{\prime 2}+b^{\prime
2})T_2-M_fM_t(aa'+bb')T_3\right)
\eeq
where $k^2=M_t^2+m_W^2-2M_tE_+$.  Note that in the standard model,
$\alpha=\alpha'=-1$, so $a=-b=2$ and $a'=b'=0$, thus the $T_2$ and
$T_3$ contributions vanish.  The contributions are
\begin{eqnarray}
T_1&= {1\over 4 w^2}\bigg[4w^2+5w-3w^3-5xw^2+3x-11xw+3w^3x -
16x^2\cr &-4x^2w+28x^3+12x^3w-16x^4-2yw^2-5yw+3w^3y\cr &-2xyw^2 
-2xy+12xyw+8x^2y-4x^2yw-8x^3y\bigg]\cr
\end{eqnarray}
where $w\equiv m^2_W$, $x=E_+$, $y=E_-$ and we have chosen units
where $M_t=1$,
\begin{eqnarray}
T_2&= {1\over 4 w^2}\bigg[w^2-w^2x-x-wx+4x^2+4wx^2-4x^3\cr
& - w^2y+wy+2xy-4wxy-4x^2y\bigg]\cr
\end{eqnarray}
and 
\beq
T_3= {1\over 4 w^2}\bigg[-11w^2-3w+26w^2x+12wx-12wx^2-4w^2y\bigg]
\eeq

The results for the branching ratio, where we have set $g_1$ and
$g_2$ to be the electroweak coupling, and $|V_{cf}|=|V_{ft}|=1$ are
given in Figure 5.  The solid curve is the contribution from $T_1$
(choosing
$a^2+b^2=1$), the dashed curve is the contribution from $T_2$
(choosing $a^{\prime 2}+b^{\prime 2}=1$) and the dotted curve is the
contribution from
$T_3$ (choosing $aa'+bb'=1$).  Note that $T_3$ is defined so that the
contribution to the square of the matrix element is negative; it is
an interference term and we have verified that the total rate is
always positive. Although we have assumed that the $V_{cf}$ and
$V_{ft}$ KM matrix elements are unity (and the cross section
obviously scales as the square of each), they are constrained
somewhat by the unitarity of the KM matrix.  These constraints are
quite weak, giving $|V_{cf}|
< 0.6$ and $V_{ft}$ is essentially
unconstrained.  (Should the value of $|V_{cf}|$ be a small as one
generally expects for mixing across two generations, then the
branching ratio would be several orders of magnitude smaller.)  Note
that the resulting branching ratio is very small, generally much less
than a part in
$10^5$.  For light masses, the branching ratio increases as one
approaches the resonance region (between $m_c+m_W$ and $m_t-m_W$), but
the current lower bound on the mass eliminates the possibility that
the heavy fermion is in that region.\footnote{D0 has
quoted\cite{dzero} a firm lower bound of
$m_Z+m_b$ on the mass; from their data it appears that a $2\sigma$
lower bound would be approximately $110-120$ GeV.}  Thus, in the case
of fermion exchange, it appears that the process \tcww  will not be
measureable, unless $|V_{cf}|$ is surprisingly large and the heavy
fermion has a mass very near its current lower bound.

For vector exchange, the diagram is given in Figure 3.  The
expression for the square of the matrix element is 
\begin{eqnarray}
|{\cal M}|^2 &={2g^2_1g^2_2(1+\alpha^2)\over w^2(s-M_V^2)^2}
\big[w^2-{1\over 4}-12w^3-19xw^2+{7x\over 4}+{7y\over 4}+12xw^3\cr\cr
&+24x^2w^2-4x^2+14x^2w+3x^3-12x^3w-19yw^2\cr\cr
&-4xw-4yw+12w^3y+24xyw^2-10xy +20xyw\cr\cr
&+17x^2y-20x^2yw-8x^3y+24y^2w^2-4y^2+14y^2w\cr\cr &+17xy^2
-20xy^2w-16x^2y^2+3y^3-12y^3w-8xy^3\big]\cr
\end{eqnarray}
where $w\equiv m^2_W$, $x=E_+$, $y=E_-$ and we have chosen units
where $M_t=1$.  Putting this expression into the branching ratio
gives the result in Figure 6.  Here, we have expressed the results
obtained by setting $g_1^2g_2^2(1+\alpha^2)=1$.  As discussed in
section 2, however, one expects $g_2$ to be two orders of magnitude
less than the weak gauge coupling, so as not to significantly mix
with the $Z$.  Thus, $g_2^2$ is at most $10^{-5}$, and thus the decay
is unmeasureable.

Part of the reason that these decay rates are so small is simply
phase space.  Any model with \tcww\ will also have \wwtc\ and one
could look for single-top/single-charm production in hadron
colliders.  Such a process will not be phase-space suppressed and
might be much larger.  We now turn to this process.

\section{\wwtc}

In order to relate a process involving $W^+W^-$ scattering to a
process involving $pp$ scattering, one needs to know the $W$ content
of the proton.  We follow the procedure of Johnson, Olness
and Tung.\cite{tung}  The cross section is
\beq
\sigma(pp\rightarrow WW)=\int\ dy_1dy_2dz_1dz_2\ f_W^d(z_1)
f_d^p(y_1)f_W^u(z_2)f_u^p(y_2)\hat{\sigma}(z_1z_2y_1y_2s)
\eeq
where, for example, $f_W^d$ is the W content of the down quark.
Changing variables to $x_1\equiv y_1z_1,\ x_2\equiv y_2z_2$ and
defining $\tau\equiv x_1x_2$, this can be written as
\beq
\sigma= \int\ {\cal L}(\tau)\hat{\sigma}(\tau s)d\tau
\eeq
where
\beq
{\cal L}(\tau)=\int_{\tau}^1\ {1\over
x_1}f_{W^-}^p(x_1)f_{W^+}^p({\tau\over x_1})\ dx_1
\eeq
Here
\beq
f_{W^-}^p(x_1)=\int_{x_1}^1 {1\over y_1}f_{W^-}^d(x_1/y_1)f_d^p(y_1)
dy_1
\eeq
with a similar expression for $f_{W^+}^p(x_2)$.  Since the structure
functions have been previously calculated (see Ref. 8), $f_{W^-}^p$
and $f_{W^+}^p$ can be determined initially, and then only the single
integral for $\sigma$ above needs to be done. It is important to note
that structure functions for longitudinal $W$'s differ from those of
transverse $W$'s, and thus the scattering cross sections must be
determined for every combination

The cross section is given by
\beq
\hat{\sigma}={|{\cal M}|^2\over 16\pi s}{1-M^2_t/s\over
\sqrt{1-4m_W^2/s}}
\eeq
and thus we only need to determine the matrix elements squared for
each process.  We will designate the transverse $W$'s as $W_L$ and
$W_R$, and the longitudinal $W$ as $W_0$.

For the scalar exchange diagram, the only nonzero contributions are
$W_LW_L$, $W_RW_R$ and $W_0W_0$.  For $W_LW_L$ and $W_RW_R$, the
square of the matrix element is
\beq
|{\cal M}|^2={g^4M_tM_cC^2_w(a^2+b^2)\over
(s-M^2_s)^2+\Gamma^2_SM^2_s}(s-M_t^2)
\eeq
and for $W_0W_0$ it is
\beq
 |{\cal M}|^2={g^4M_tM_cC^2_w(a^2+b^2)\over 
(s-M^2_s)^2+\Gamma^2_SM^2_s}(s-M_t^2)\left({s\over 2m_W^2}-1\right)^2
\eeq

For the fermion exchange diagram, there are many more spin
combinations. The full expressions are given in the Appendix.
For the vector exchange diagram, the only non-vanishing contributions
are $W_LW_L$ (which is identical to $W_RW_R$), $W_L^+W_0^-$ (which
is identical to $W_R^-W_0^+$) and $W_0W_0$.  For $W_LW_L$, we find
\beq
{|\cal M|}^2={g_1^2g_2^2\over
(s-M_V^2)^2}{(1+\alpha^2)(s-M_t^2)(s-4m_W^2)(M_t^2+2s)\over
 3s}
\eeq 
For $W_L^+W_0^-$, we have the same expression multiplied by
$s/2m^2_W$,
and finally, for $W_0W_0$, we have the same expression multiplied by
$(s/2m^2_W+1)^2$.

\section{Results and Discussion}

The results are given in Fig. 7.  Results are given for the
LHC ($\sqrt{s}=14$ TeV).  
We see that the cross sections are all quite small, never exceeding
$10^{-3}$ picobarns.   Note that we have assumed that the scalar
boson flavor changing coupling to the top and charm is the geometric
mean of the top and charm Yukawa couplings; if it were the top
Yukawa coupling, the rate could be $0.1$ picobarns. 
Note also that we have scaled the cross-section in the case of
vector exchange by $C_w^2=10^{-4}$, corresponding to the 
model-dependent bound on mixing with the $Z$; should mixing with
the $Z$ be suppressed, this coupling could be larger.

Could such a cross section lead to measureable single-top/single-charm
production?  After all, a $10^{-3}$ picobarn
cross section, with an integrated luminosity of $1000$ inverse
femtobarns (possible at the LHC after several years of running), will
lead to $1000$ events.  However, there will be a background from
single-top/single-bottom production.  The cross section for
$pp\rightarrow q\bar{q}\rightarrow t\bar{b}$, through s-channel
W-exchange, is $10$ picobarns at the LHC\cite{scott} (the cross
section for single top production through gluon-W fusion is larger,
but the $b$ is much softer and appropriate kinematic cuts can suppress
this signal).  Thus, one would have to be able to distinguish  a
$\bar{c}$ in a sample of $10^4\ $ $\bar{b}$'s.  Alhough kinematic cuts
and good vertex detection would help, and the expected
single-top/single-bottom cross section can be calculated to a few
percent accuracy,  it is hard to imagine that an unambiguous signal
could be detected.  It may not be hopeless in the scalar exchange case
if the coupling of the scalar was unexpectedly large; the rate could
then be as high as $0.1$ picobarns, and one would get a
$\bar{c}$ for every hundred $\bar{b}$'s.  In this case, the signal
possibly could be extracted.

In view of our statements in the introduction, the fact that \wwtc\
appears to be too small (for the most natural values of couplings) to
be detected, while
\tcww\ can, in some cases, be detected, may be surprising.  After all,
the latter does have a significant phase space suppression, while the
former does not.  However, the latter also has the possibility of a
resonance for some mass range, while this resonance is absent
in the hadronic collisions.

We conclude that the process \wwtc\ is probably undetectable at 
the LHC. The only hope would occur if the exchanged
scalar boson had a surprisingly large coupling.   The process
\tcww\ may be detectable in two cases:  (a) With scalar exchange and
the scalar mass between $2m_W$ and approximately $200$ GeV and (b)
with a heavy fermion with a mass very near the current lower bound
and with an unexpectedly high mixing with the second and third
generations.

\vspace*{.25 in}

This
research was supported in part by the U.S. 
DOE contract DC-AC05-84ER40150
(Jefferson Lab).

\section{Appendix}

Here, we give the results for the fermion exchange diagram in terms of
the scattering angle. Integrating over this angle is straightforward. 
We define
$r\equiv \cos^2{\theta\over 2}$ and $\bar{r}\equiv \sin^2{\theta\over
2}$, and also define $K_o\equiv g^4|V_{tf}V_{cf}^*|^2/(t-m_F^2)^2$.
Results are given for all possible combinations of $\alpha$ and
$\alpha'$ equal to $\pm 1$; other cases may be derived from those.

For $W_LW_L$ scattering, the square of the matrix element is
$K_os^{-2}(s-M_t^2)A$, where for $(\alpha,\alpha')=(1,1)$,
$A=(s-M^2_t)^2r\bar{r}^2(M^2_tr+s\bar{r})$; for
$(\alpha,\alpha')=(1,-1)$, $A=m_F^2\bar{r}(M^2_tr+s\bar{r})$; for
$(\alpha,\alpha')=(-1,1)$, $A$ is the same with $r\leftrightarrow
\bar{r}$, and for $(\alpha,\alpha')=(-1,-1)$,
$A=\bar{r}(M_t^2r+s\bar{r})(M_t^2\bar{r}+sr)^2$.  For $W_RW_R$
scattering, the result is the same

For $W_R^+W_L^-$ scattering, only the $\alpha=\alpha'$ contributions
are nonzero.  For $(\alpha,\alpha')=(1,1)$, the square of the matrix
element is $K_os^{-2}(s-M_t^2)^3r^2\bar{r}(\bar{r}M^2_t+rs)$ and for
$(\alpha,\alpha')=(-1,-1)$ it is the same with $r\leftrightarrow
\bar{r}$.  For $W_L^+W_R^-$, the result is also the same.

For $W_0^+W_L^-$, the square of the matrix element is ${1\over 2}K_o
s^{-2}(s-M_t^2)m_W^{-2}(s\bar{r}+M^2_tr)A$, where for
$(\alpha,\alpha')=(1,1)$,
$A=s(s-M^2_t)^2\bar{r}r^2$; for $(\alpha,\alpha')=(-1,+1)$,
$A=s^2rM_F^2$ and for $(\alpha,\alpha')=(-1,-1)$,
$A=\bar{r}(s\bar{r}+M_t^2r)^2$.  For the remaining case,
$(\alpha,\alpha')=(1,-1)$, the result vanishes.  Again, for
$W_0^+W_R^-$, the result is the same.

For $W_0^-W_R^+$, the square of the matrix element is ${1\over 2}K_o
s^{-2}(s-M_t^2)m_W^{-2}(sr+M^2_t\bar{r})A$, where for
$(\alpha,\alpha')=(1,1)$,
$A=s(s-M^2_t)^2r^3$; for $(\alpha,\alpha')=(1,-1)$,
$A=s^2\bar{r}M_F^2$ and for $(\alpha,\alpha')=(-1,-1)$,
$A=\bar{r}^2r$.  For the remaining case,
$(\alpha,\alpha')=(-1,1)$, the result vanishes.  Again, for
$W_0^-W_L^+$, the result is the same.

Finally, for $W_0W_0$, the square of the matrix element is
${1\over 4}K_om_W^{-4}(s-M_t^2)(s\bar{r}+M_t^2r)A$.  For
$\alpha=\alpha'$, $A=(s-M_t^2)^2r\bar{r}^2$ and for
$\alpha=-\alpha'$, $A=sm_F^2\bar{r}$.

\newpage
\centerline{\bf FIGURE CAPTIONS}
\parindent=0pt
\vskip 2cm

{\bf Fig. 1} This process gives \tcww\ in the standard model, where
$F$ is any $Q=-1/3$ quark.  In generic non-standard models, $F$ is a
heavy quark.  We define the upper vertex to be $ig_1
\gamma_\mu(1+\alpha\gamma_5)/2\sqrt{2}$ and the lower to be
 $ig_2
\gamma_\nu(1+\alpha'\gamma_5)/2\sqrt{2}$.

{\bf Fig. 2} Scalar exchange contribution to \tcww\ in a multi-scalar
model.  The Yukawa coupling is defined to be
$-i(a+b\gamma_5)g\sqrt{m_cm_t}/\sqrt{2}m_W$ and the
scalar-vector-vector coupling is defined to be $igm_Wg_{\mu\nu}C_w$.

{\bf Fig. 3} Vector exchange contribution to \tcww.  The Yukawa
coupling is defined to be $ig_2\gamma_\rho (1+\alpha\gamma_5)/2$
and the triple-vector coupling is defined to be
$g_1(g_{\mu\nu}(k-p)_\rho+
g_{\nu\rho}(p-q)_\mu+g_{\rho\mu}(q-k)_\nu)$, where all momenta are
defined to flow outward from the vertex.

{\bf Fig. 4} Branching ratio for \tcww\ due to non-standard model
scalar exchange, using the vertices from Figure 2.  The rate is
proportional to $C_w^2(a^2+b^2)$, which we have set equal to unity
in the figure.

{\bf Fig. 5}  Branching ratio for \tcww\ due to heavy fermion
exchange.  The solid (dashed, dotted) curve is the contribution from
$T_1$ ($T_2$,$T_3$), defined in Eq. 3.  Here, we have set $g_1$ and
$g_2$ equal to the electroweak gauge coupling, and set $|V_{cf}|$ and
$|V_{ft}|$ equal to unity; the branching ratio scales with these
quantities as shown in Eq. 3.

{\bf Fig. 6} Branching ratio for \tcww\ due to non-standard model
vector exchange, using the vertices from Figure 3.  In plotting the
above, we have set $g_1^2g_2^2(1+\alpha^2)$ equal to unity. 
As discussed in the text, in most models limits from mixing with
the standard model $Z$ force $g_1$ to be roughly two orders of
magnitude smaller than the gauge coupling; the branching ratio in
such a case will be more than four orders of magnitude smaller than
shown in the figure.

{\bf Fig. 7} 
Cross section for \wwtc\ at the LHC ($
\sqrt{s}=14$ TeV).
For vector
exchange indicated
by the dashed line, the value of $C_w^2$ has been taken to be $10^{-4}$, 
which is the maximum value expected from limits on $Z$ mixing.
The solid line corresponds to scalar exchange, 
The other lines indicate the cross section for fermion exchange. The
labels $\{LL$, $LR$, $RL$, $RR\}$ refere to the cases where 
$(\alpha, \alpha^\prime)$ are $\{(-1,-1)$, $(-1,+1)$, $(+1,-1)$, 
$(+1,+1)\}$ respectively.
\newpage
\begin{figure}[h]
\epsfysize 6.0in
\mbox{\epsfbox{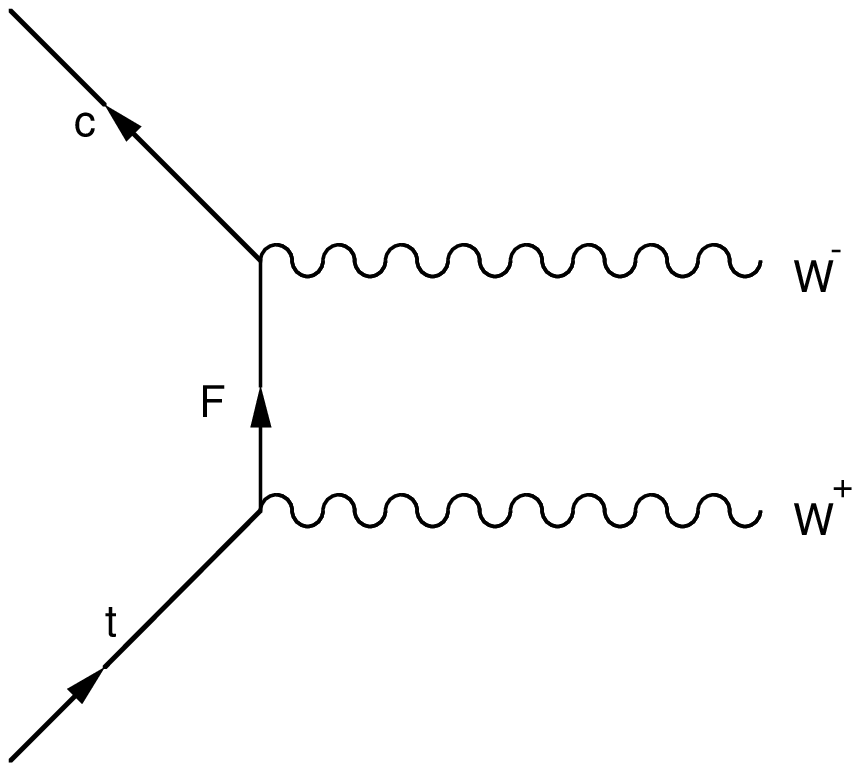}}
\end{figure}
\newpage
\begin{figure}[h]
\epsfysize 6.0in
\mbox{\epsfbox{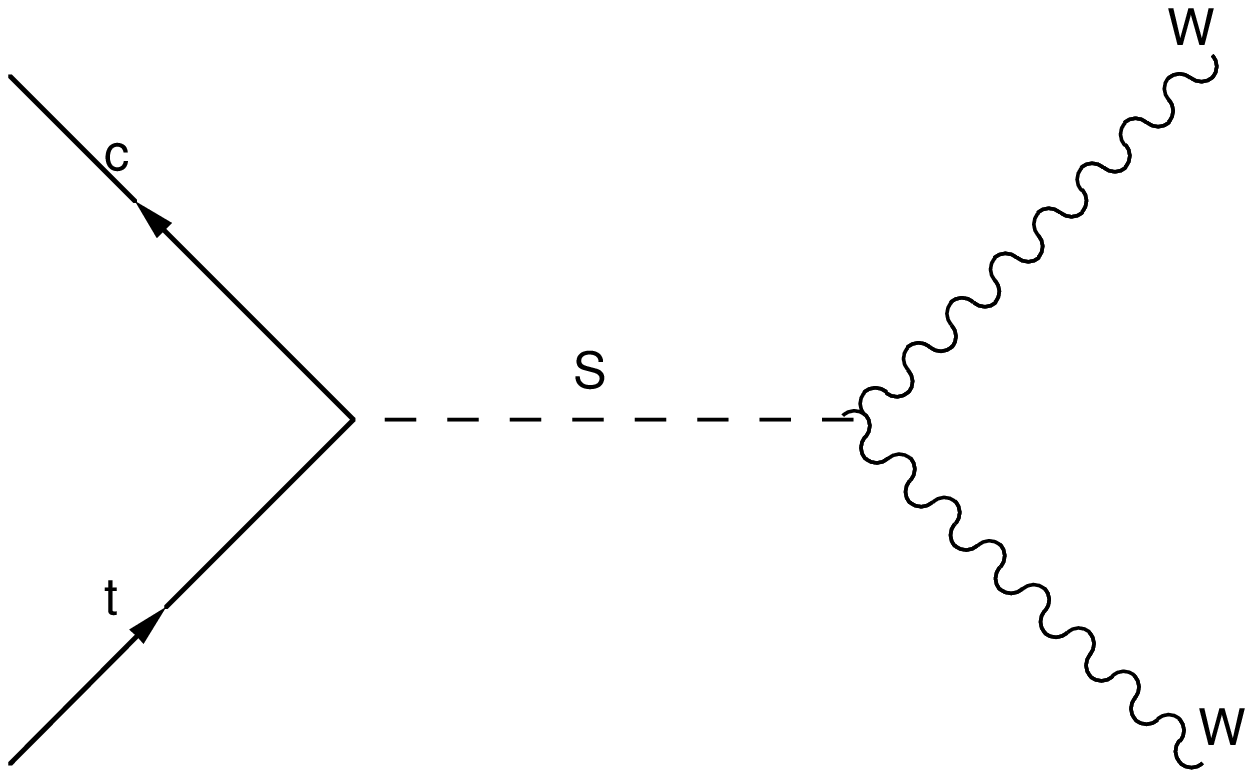}}
\end{figure}
\newpage
\begin{figure}[h]
\epsfysize 6.0in
\mbox{\epsfbox{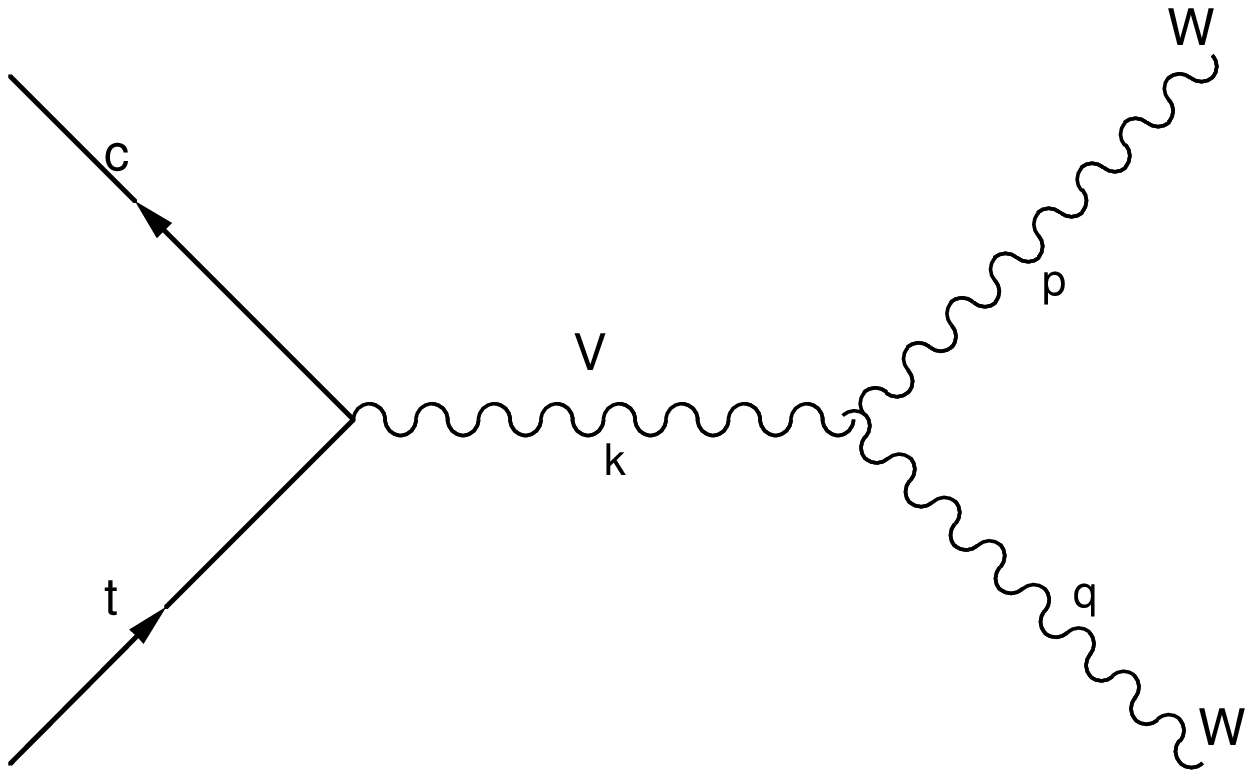}}
\end{figure}
\newpage
\begin{figure}[h]
\hspace*{0 in}
\epsfysize 7.00in
\mbox{\epsfbox{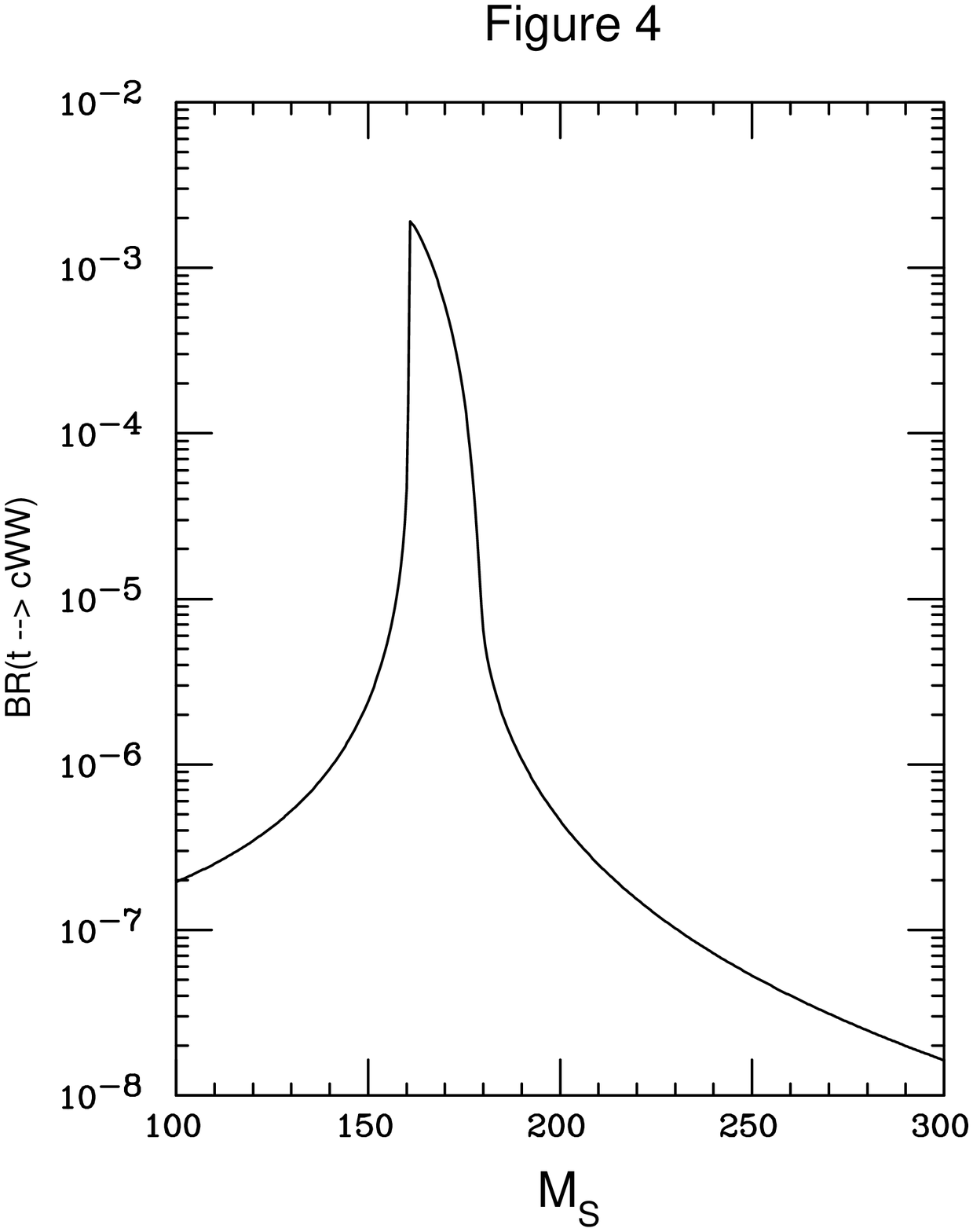}}
\end{figure}
\newpage
\begin{figure}[h]
\hspace*{0 in}
\epsfysize 7.00in
\mbox{\epsfbox{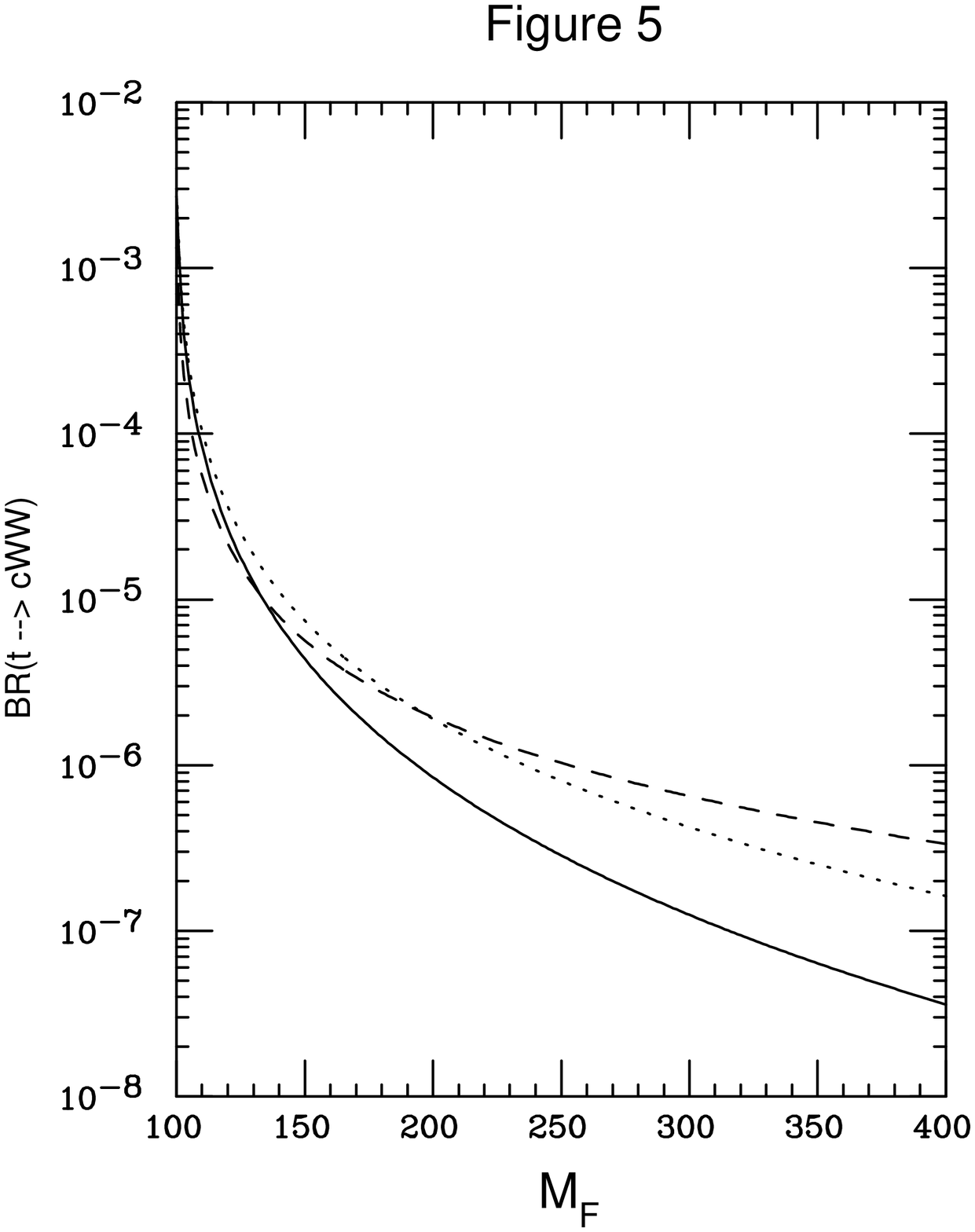}}
\end{figure}
\newpage
\begin{figure}[h]
\hspace*{0 in}
\epsfysize 7.00in
\mbox{\epsfbox{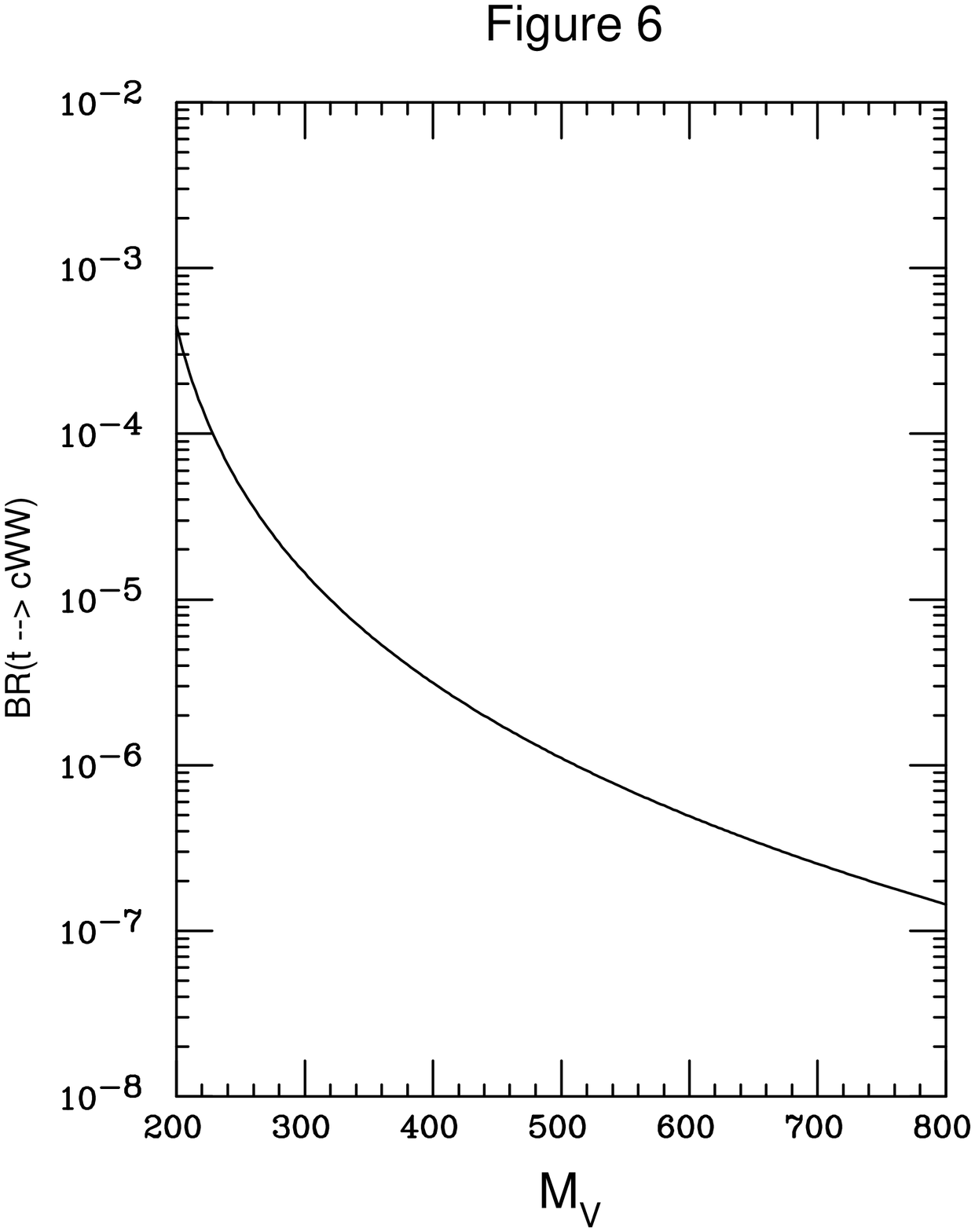}}
\end{figure}
\newpage
\begin{figure}[h]
\hspace*{0 in}
\epsfysize 7.00in
\mbox{\epsfbox{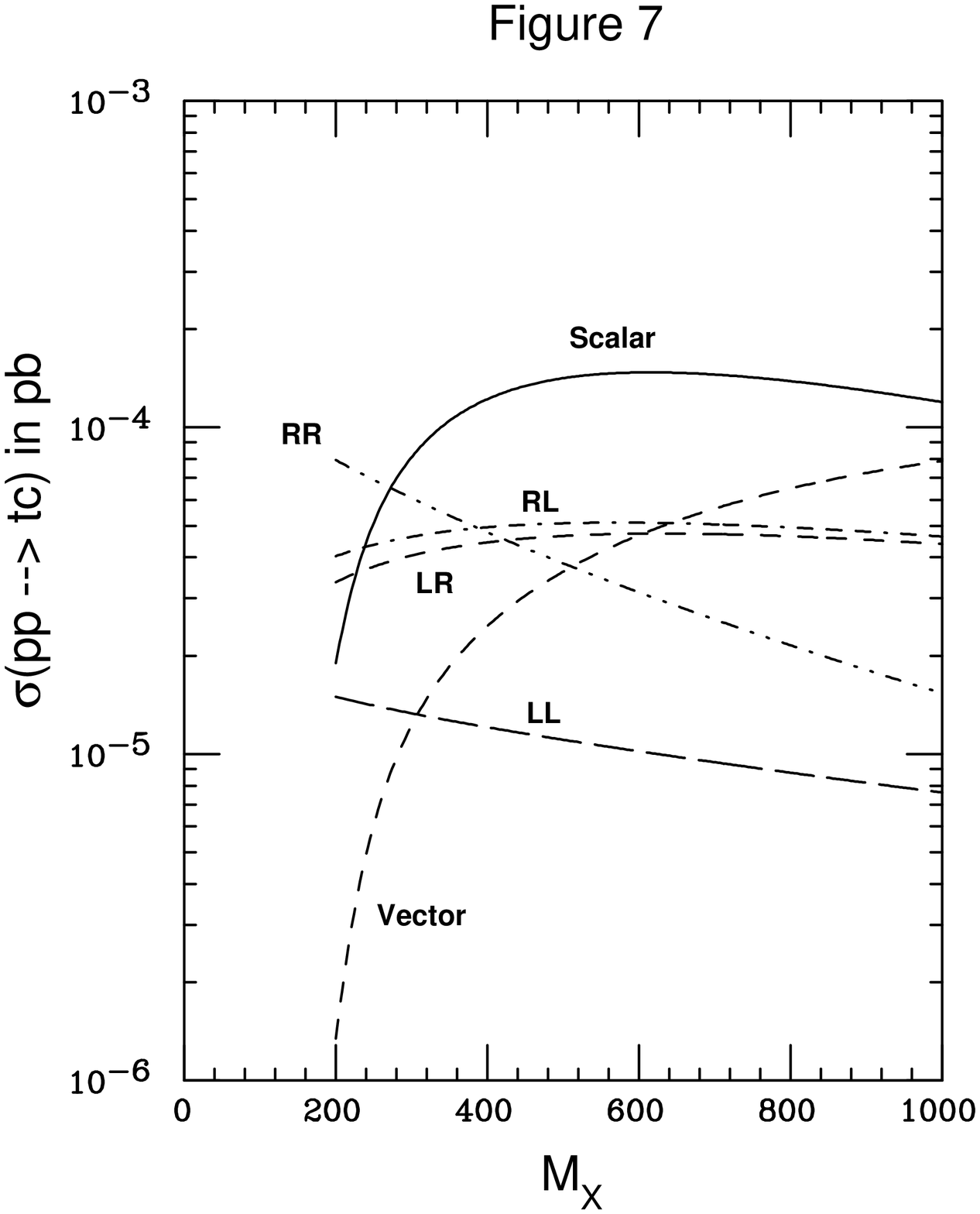}}
\end{figure}
\end{document}